\documentclass[12pt]{article}
\usepackage{amsfonts}
\usepackage{amsmath}
\usepackage{amssymb}
\usepackage{graphicx}
\usepackage{color}
\usepackage[all, knot]{xy}
\usepackage{tikz}

\usepackage{ulem}

\usepackage[utf8]{inputenc}
\usepackage{epstopdf}
\usepackage[footnotesize]{caption}
\usepackage{amsthm}
\usepackage{enumitem}
\usepackage{mathrsfs}

\usepackage[margin=3cm]{geometry}

\def \be {\begin{equation}}
\def \ee {\end{equation}}
\def \bea {\begin{eqnarray}}
\def \eea {\end{eqnarray}}
\def \nn {\nonumber}

\def \rr {\raise.35ex\hbox{\small $\prime$}\kern-.17em{\mbox{\large $\imath$}}}

\def \dels {\partial\kern-.6em /\kern.1em}
\def \As {{A\kern-.5em / \kern.5em}}
\def \Ds {D\kern-.7em / \kern.5em}

\def \ks {k\kern-.5em /}
\def \ls {l\kern-.5em /}

\newcommand{\ci}[1]{}

\newcommand{\ba}{\begin{eqnarray}}
\newcommand{\ea}{\end{eqnarray}}
\newcommand{\bal}{\begin{align}}
\newcommand{\eal}{\end{align}}
\newcommand{\bay}[1]{\left(\begin{array}{#1}}
\newcommand{\eay}{\end{array}\right)}

\def\xe{{\epsilon}}

\newcommand{\hide}[1]{}

\newlist{axioms}{enumerate}{2}
\setlist[axioms,1]{label=\textbf{A\arabic{axiomsi}.}, ref=A\arabic{axiomsi}}
\setlist[axioms,2]{label=\textbf{A\arabic{axiomsi}\rlap{\myEnumCounter{axiomsii}}.},%
                   ref=A\arabic{axiomsi}\myEnumCounter{axiomsii},%
                   align=parleft,%
                   leftmargin=0em,%
                   itemsep=1.4ex,%
                   before={\stepcounter{axiomsi}}}

  \usetikzlibrary{decorations.markings}

\begin{document}

\begin{titlepage}
\begin{center}

\textbf{\LARGE
Emergence of Kinematic Space from\\ 
Quantum Modular Geometric Tensor
\vskip.3cm
}
\vskip .5in
{\large
Xing Huang$^{a,b,c}$ \footnote{e-mail address: xingavatar@gmail.com} and
Chen-Te Ma$^{d,e,f,g,h}$ \footnote{e-mail address: yefgst@gmail.com}
\\
\vskip 1mm
}
{\sl
$^a$
Institute of Modern Physics, Northwest University, Xi'an 710069, China.
\\
$^b$
NSFC-SPTP Peng Huanwu Center for Fundamental Theory, Xi'an 710127, China.
\\
$^c$
Shaanxi Key Laboratory for Theoretical Physics Frontiers, Xi'an 710069, China. 
\\
$^d$
Asia Pacific Center for Theoretical Physics,\\
Pohang University of Science and Technology, 
Pohang 37673, Gyeongsangbuk-do, South Korea. 
\\
$^e$
Guangdong Provincial Key Laboratory of Nuclear Science,\\
 Institute of Quantum Matter,
South China Normal University, Guangzhou 510006, Guangdong, China. 
\\ 
$^f$ 
School of Physics and Telecommunication Engineering,\\
South China Normal University, Guangzhou 510006, Guangdong, China. 
\\ 
$^g$
Guangdong-Hong Kong Joint Laboratory of Quantum Matter,\\
 Southern Nuclear Science Computing Center, 
South China Normal University, 
Guangzhou 510006, Guangdong, China. 
\\
$^h$ 
The Laboratory for Quantum Gravity and Strings,\\
 Department of Mathematics and Applied Mathematics, 
University of Cape Town,
 Private Bag, Rondebosch 7700, South Africa.
}\\
\vskip 1mm
\vspace{40pt}
\end{center}

\newpage
\begin{abstract} 
We generalize the Quantum Geometric Tensor by replacing a Hamiltonian with a modular Hamiltonian. 
The symmetric part of the Quantum Geometric Tensor provides a Fubini-Study metric, and its anti-symmetric sector gives a Berry curvature. 
Now the generalization or Quantum Modular Geometric Tensor gives a Kinematic Space and a modular Berry curvature. 
Here we demonstrate the emergence by focusing on a spherical entangling surface. 
We also use the result of the identity Virasoro block to relate the connected correlator of two Wilson lines to the two-point function of a modular Hamiltonian. 
This result realizes a novel holographic entanglement formula for two intervals of a general separation. 
 This formula does not only hold for a classical gravity sector but also Quantum Gravity. 
 The formula also provides a new Quantum Information interpretation to the connected correlators of Wilson lines as the mutual information.
 Our study provides an opportunity to explore Quantum Kinematic Space through Quantum Modular Geometric Tensor and hence go beyond symmetry. 
\end{abstract}
\end{titlepage}

\section{Introduction}
\label{sec:1}
\noindent
The {\it Emergent Spacetime} proposes that space and time are not fundamental but emergent (at a macroscopic scale).  
The successful approach is based on this idea, like ($d+1$)-dimensional Anti-de Sitter/$d$-dimensional Conformal Field Theory (AdS$_{d+1}$/CFT$_d$) correspondence, which avoids the difficulty of non-renormalizability in Quantum Gravity.  
Therefore, one can consider Quantum Field Theory (QFT) on a flat spacetime for an alternative approach of Quantum Gravity. 
The correspondence transforms the problem into the building of a dictionary of bulk gravity theory and QFT. 
The classical gravity theory and CFT \cite{Dolan:2011dv} are both calculable. 
Therefore, it is easier than non-renormalizability. 
Recently, people were interested in using a picture of Quantum Information to interpret the holographic principle. 
It was motivated by the emergence of an AdS$_{d+1}$ minimum surface from CFT$_d$ entanglement entropy. 
Indeed, people did not understand how a bulk gravity emerges from physical degrees of freedom in QFT \cite{Perlmutter:2013gua, Casini:2011kv}. 
The holographic formula implies that emergent spacetime comes from the degrees of Quantum Entanglement.
\\

\noindent
The quantum correction in the bulk gravity generates a {\it conformal anomaly}. 
The conformal symmetry is {\it not} enough in the AdS/CFT correspondence. 
Therefore, it is hard to understand why the correspondence still holds in a quantum regime. 
The gauge formulation of 3d Einstein gravity theory has an exact correspondence from a non-conformal field theory. 
A bulk Wilson line operator \cite{Fitzpatrick:2016mtp} (or a quantum minimum-surface) is dual to entanglement entropy for a single interval \cite{Huang:2019nfm, Huang:2020tjl}. 
Therefore, at the quantum level, we still have a geometry that emerged from entanglement entropy. 
It is a non-trivial realization of the holographic entanglement formula beyond a classical gravity sector. 
One can also calculate a connected two-point function of modular Hamiltonian or mutual information of two intervals for detecting quantum bulk gravity theory \cite{Faulkner:2013ana}. 
It is hard to know whether the emergence can happen in a quantum regime because it is hard to compute.      
\\

\noindent 
It is convenient to calculate $H_{\mathrm{mod}}$ for the two regions are separated by a spherical entangling surface:
\bea
H_{\mathrm{mod}}\equiv-\ln\rho_A, \ \rho_A\equiv\mathrm{Tr}_B\rho_{AB}, 
\eea
where $\rho_{AB}$ is a density matrix with a Hilbert space $H_A\otimes H_B$, $\rho_A$ is a reduced density matrix of a region $A$, and $\mathrm{Tr}_B$ is a partial trace operation acting on a region $B$.  
People used the operator product expansion (OPE) to organize the CFT data and extract the {\it OPE block} $B_l^{jk}$ \cite{Czech:2016xec}, 
\bea
{\cal O}_j(x_1){\cal O}_k(x_2)\equiv|x_1-x_2|^{-\Delta_j-\Delta_k}\sum_lC_{jkl}B_l^{jk}(x_1, x_2), 
\eea  
where $\Delta_j$ and $\Delta_k$ are the conformal dimensions of ${\cal O}_j$ and ${\cal O}_k$, respectively, and $C_{jkl}$ is an OPE coefficient. 
The $B_l^{jk}$ is proportional to a modular Hamiltonian, and it satisfies a Laplacian equation on a 2$d$-dimensional {\it Kinematic Space} (KS) \cite{Czech:2016xec, deBoer:2016pqk}
\bea
ds^2\equiv g_{\mu\nu}dx^{\mu}dy^{\nu}
=\alpha\frac{4}{(x-y)^2}
\bigg(-\eta_{\mu\nu}+\frac{2(x_{\mu}-y_{\mu})(x_{\nu}-y_{\nu})}{(x-y)^2}\bigg)dx^{\mu}dy^{\nu}, 
\eea
where 
\bea
(x-y)^2\equiv \eta^{\mu\nu}(x_{\mu}-y_{\mu})(x_{\nu}-y_{\nu}); \qquad 
\eta_{\mu\nu}\equiv\mathrm{diag}(-1, 1, \cdots, 1).
\eea
The $\alpha$ is an arbitrary constant.  
We label the indices of boundary spacetime as $\mu=0, 1, \cdots, d-1$. 
For $d=2$, the metric of KS is given by a second-order derivative of geodesic length (kinematic measure) \cite{Czech:2015qta}. 
In a higher-dimensional CFT, the KS metric remains the second derivative of a similar logarithmic function \cite{deBoer:2016pqk}, whose connection with the minimum surface is unclear. 
Moreover, the definition of KS for a general entangling surface is still an open question. 
\\

\noindent 
When a spin-1/2 electron adiabatically follows a magnetic field, a gauge field provides a Berry curvature to affect the motion. 
Nowadays, people discovered that this phenomenon is just a holonomy. 
The difference of a quantum state is by a U(1) phase. 
One can geometrize usual Quantum Mechanics through an emergent gauge potential (defines a covariant derivative). 
Because a quantum distance between different states should be invariant for a different holonomy, the Hilbert space has a U(1) quotient. 
Requiring the U(1) gauge invariance on Fubini-Study metric and Berry curvature can formulate the combination, {\it Quantum Geometric Tensor} (QGT) \cite{Provost:1980nc}. 
The QGT gives a beautiful {\it unification} of a metric and curvature. 
Recently, one provided a {\it modular} extension to a Berry curvature \cite{Czech:2017zfq, Czech:2019vih}, similar to replacing a Hamiltonian by $H_{\mathrm{mod}}$. 
One could also show the equivalence between the {\it modular Berry geometry} and {\it Riemann geometry} on KS \cite{Huang:2019wzc, Huang:2020cye}. 
Therefore, we expect to obtain KS and modular Berry curvature (MBC) from {\it Quantum Modular Geometric Tensor} \cite{Provost:1980nc}
\bea
g_{jk}^{(n)}(\vec{\lambda})\equiv
\langle \partial_jn(\vec{\lambda})|\partial_kn(\vec{\lambda})\rangle
-\langle \partial_jn(\vec{\lambda})|n(\vec{\lambda})\rangle\langle n(\vec{\lambda})|\partial_kn(\vec{\lambda})\rangle, 
\eea 
where $|n(\vec{\lambda})\rangle$ is an eigenstate of $H_{\mathrm{mod}}$, and the derivative of the eigenstate is given by
\bea
|\partial_jn(\vec{\lambda})\rangle\equiv \frac{\partial}{\partial\lambda_j}|n(\vec{\lambda})\rangle. 
\eea 
The central question that we would like to address in this letter is the following: {\it What is the relation between KS and QMGT?}
\\

\noindent
In this letter, we realize KS$_{2d}$/CFT$_d$ correspondence through QMGT. 
Our result implies that KS and MBC are given by the symmetric and anti-symmetric parts of QMGT, respectively. 
On a CFT$_2$ vacuum state, we obtain a novel entanglement formula for two intervals of a general separation. 
This formula provides a new interpretation of connected Wilson lines from mutual information.  

\section{QMGT and Emergent Geometry}
\label{sec:2} 
\begin{figure}
\begin{center}
\includegraphics[width=1.\textwidth]{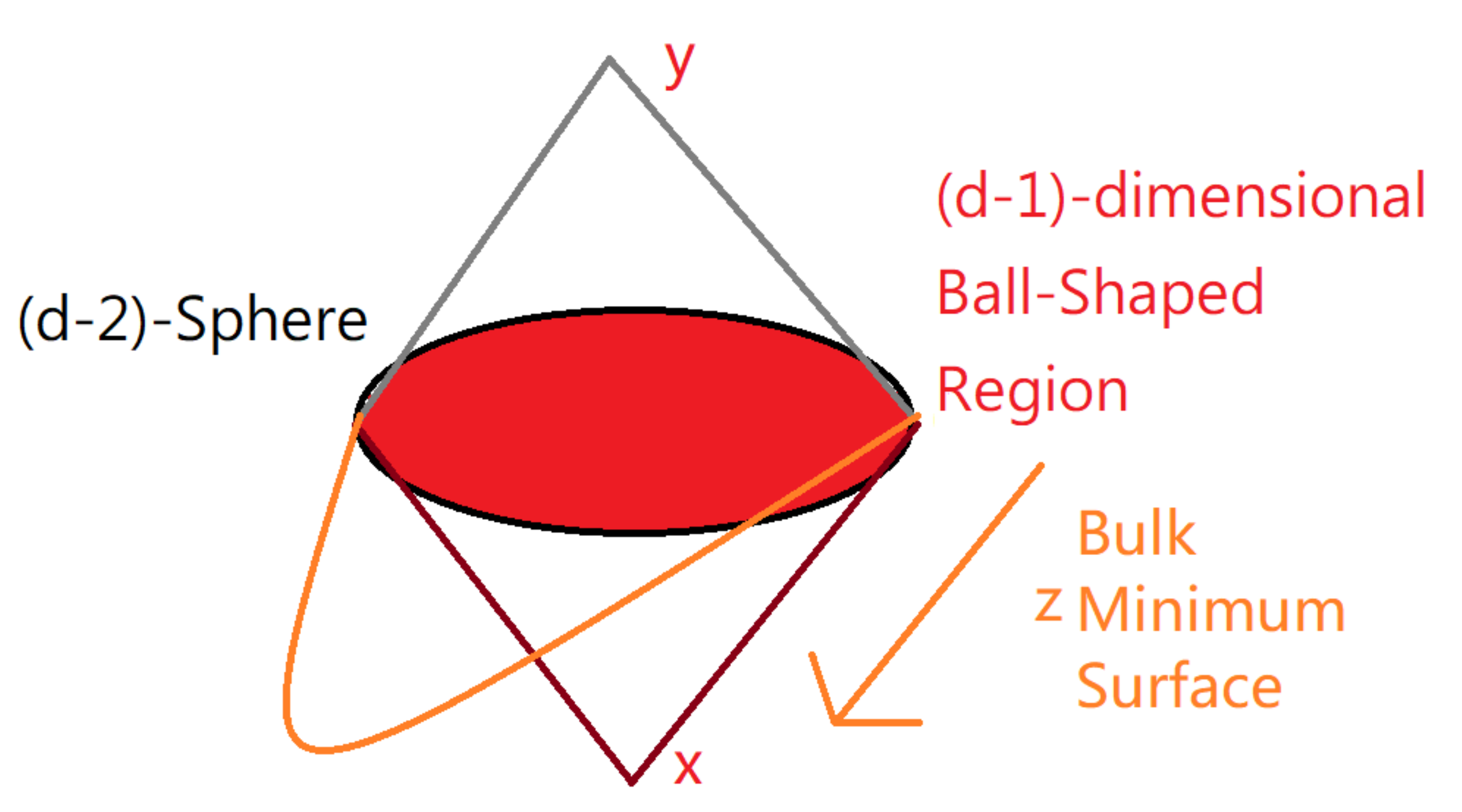}
\end{center}
\caption{A casual diamond in $d=3$. 
The ($d-2$)-dimensional spherical region is specified by a pair of time-like separated points $x^{\mu}$ and $y^{\mu}$. }
\label{BR.pdf}
\end{figure}

\noindent 
We first calculate QMGT in CFTs for a modular Hamiltonian with a spherical entangling surface, which satisfies the following algebra: 
 \bea
 \lbrack H_{\mathrm{mod}}, H_{\mathrm{mod}}\rbrack&=&0; 
 \nn\\
 \lbrack H_{\mathrm{mod}}, \partial_{\nu, x} H_{\mathrm{mod}}\rbrack&=&-2\pi i\partial_{\nu, x}H_{\mathrm{mod}};
 \nn\\
 \lbrack H_{\mathrm{mod}}, \partial_{\nu, y} H_{\mathrm{mod}}\rbrack&=&2\pi i\partial_{\nu, y}H_{\mathrm{mod}},  \label{algebradeform}
 \eea 
 where 
 \bea
 \partial_{\nu, x}\equiv\frac{\partial}{\partial x^{\nu}}, \ 
 \partial_{\nu, y}\equiv\frac{\partial}{\partial y^{\nu}}. 
 \eea
 The $x$ and $y$ are the tips of a causal diamond as in Fig. \ref{BR.pdf} \cite{deBoer:2016pqk}. 
A direct calculation shows that the anti-symmetric part of QMGT is MBC. 
We then use CFT$_2$ and CFT$_3$ to demonstrate our method for extracting the metric of KS \cite{Czech:2016xec, deBoer:2016pqk} from the symmetric part of QMGT and present the results in general dimensions. 

\subsection{QMGT in CFTs} 
\noindent 
Because $H_{\mathrm{mod}}$ is diagonal for the eigenstate $|n\rangle$, we obtain that: 
\bea
0&=&\partial_j\big(\langle m|H_{\mathrm{mod}}|n\rangle\big)
=E_{\mathrm{mod}}^{(n)}\langle\partial_j m|n\rangle 
+E_{\mathrm{mod}}^{(m)}\langle m|\partial_j n\rangle
+\langle m|\partial_jH_{\mathrm{mod}}|n\rangle, \ m\neq n; 
\nn\\ 
\langle m|\partial_j n\rangle&=&\frac{\langle m|\partial_j H_{\mathrm{mod}}|n\rangle}
{E_{\mathrm{mod}}^{(n)}-E_{\mathrm{mod}}^{(m)}}, 
\eea
where $E_{\mathrm{mod}}^{(n)}$ is an eigenvalue of $H_{\mathrm{mod}}$ for $|n\rangle$. 
Therefore, the QMGT becomes 
 \bea
 g_{jk}^{(n)}=\sum_{m\neq n}\frac{\langle n|\partial_j H_{\mathrm{mod}}|m\rangle\langle m|\partial_kH_{\mathrm{mod}}|n\rangle}{(E_{\mathrm{mod}}^{(n)}-E_{\mathrm{mod}}^{(m)})^2}.
 \eea 
 \\ 
 
 \noindent 
 Now we show $E^{(n)}_{\mathrm{mod}}=E^{(m)}_{\mathrm{mod}}\pm 2\pi i$ when $\langle m|\partial_{\nu, x; y}H_{\mathrm{mod}}|n\rangle\neq 0$ as in the following: 
 \bea
 \langle m|\lbrack H_{\mathrm{mod}}, \partial_{\nu, x; y}H_{\mathrm{mod}}\rbrack|n\rangle
 &=&(E_{\mathrm{mod}}^{(m)}-E_{\mathrm{mod}}^{(n)})
 \langle m|\partial_{\nu, x; y}H_{\mathrm{mod}}|n\rangle 
 \nn\\
 &=&\mp 2\pi i\langle m|\partial_{\nu, x; y}H_{\mathrm{mod}}|n\rangle. 
 \eea
 We use Eq. \eqref{algebradeform} in the second equality. 
 This result implies $\langle n|\partial_{\nu, x; y}H_{\mathrm{mod}}|n\rangle=0$.  
 Hence the QMGT is the same as a two-point function of $H_{\mathrm{mod}}$ for the spherical case 
 \bea
 g_{jk}^{(n)}=
 \sum_{m\neq n}\frac{\langle n|\partial_j H_{\mathrm{mod}}|m\rangle\langle m|\partial_kH_{\mathrm{mod}}|n\rangle}{(2\pi i)^2}
 =\frac{\langle n|\partial_j H_{\mathrm{mod}}\partial_kH_{\mathrm{mod}}|n\rangle}{(2\pi i)^2}.
 \eea 
 When $\partial_j=\partial_{\mu, x}$ ($\partial_{\mu, y}$) and $\partial_k=\partial_{\nu, x}$ ($\partial_{\nu, y}$), we cannot find $|m\rangle$ to satisfy  $E^{(n)}_{\mathrm{mod}}-E^{(m)}_{\mathrm{mod}}=\pm 2\pi i$. 
 Therefore, it implies $\langle m|\partial_{\nu, x; y}H_{\mathrm{mod}}|n\rangle=0$. 
 Hence we only need to consider $\partial_j=\partial_{\mu, x}$ ($\partial_{\mu, y}$) and $\partial_k=\partial_{\nu, y}$ ($\partial_{\nu, x}$).

\subsection{KS and MBC} 
\noindent
The anti-symmetric part of QMGT for $\partial_{\mu, x}$ and $\partial_{\mu, y}$ provides the only non-trivial component 
\bea
{\cal R}_{\mu,x; \nu, y}=-\frac{1}{4\pi^2}\lbrack\partial_{\mu, x} H_{\mathrm{mod}}, \partial_{\nu, y}H_{\mathrm{mod}}\rbrack,   
\eea
which is consistent with MBC \cite{Czech:2017zfq}. 
Because MBC is an operator, it is unnecessary to calculate the expectation value by a regularization. 
\\

\noindent
Now we calculate the symmetric part of QMGT. 
In general dimensions, the 2-point function of stress tensor shows that 
\bea
\langle T_{\mu\nu}(x)T_{\rho\sigma}(0)\rangle=\frac{C_TI_{\mu\nu, \rho\sigma}}{x^{2d}}, 
\eea
where 
\bea
I_{00, 00}=1-\frac{1}{d}. 
\eea
The $C_T$ is related to the $c$-anomaly $c_d$ as 
\bea
C_T=\frac{(d+1)!}{(d-1)\pi^d}c_d. 
\eea
In the holographic CFT, the $c_d$ equals to the $a$-anomaly $a_d$ \cite{Perlmutter:2013gua}. \\

\noindent 
For a spherical entangling surface and $d>1$, the modular Hamiltonian is \cite{Casini:2011kv}
\bea
H_{\mathrm{mod}}=2\pi\int_A d^{d-1}x^{\prime}\ \frac{R^2-|\vec{x}-\vec{x}^{\prime}|^2}{2R}T_{00}(\vec{x}^{\prime}),
\eea
where $T_{00}$ is the (00)-component of a stress tensor, and $R$ is the radius of the $(d-1)$-dimensional sphere centered at $x$. 
When $d=1$, we can the OPE block to define the modular Hamiltonian, but the integration variable becomes time \cite{Huang:2019wzc, Huang:2020cye}. 
When the radius $R$ approaches zero, the modular Hamiltonian becomes: 
\bea
H_{\mathrm{mod}}\rightarrow 
\frac{4\pi^{\frac{d-1}{2}+1}R^d}{(d^2-1)\Gamma(\frac{d-1}{2})}T_{00}
\equiv A_H\big(-(x^{\mu}-y^{\mu})(x_{\mu}-y_{\mu})\big)^{\frac{d}{2}} T_{00}
=2^dR^d A_H T_{00}
, 
\eea
where 
\bea
A_H=\frac{4\pi^{\frac{d-1}{2}+1}}{2^d(d^2-1)\Gamma(\frac{d-1}{2})}. 
\eea 
Hence the 2-point function of $H_{\mathrm{mod}}$ is different
from the conformal block with asymptotic behavior $z^d$ in $z\to 0$ (the definition is in Eq. \eqref{uv}) by following factor: 
\bea
C_1\equiv A_H^2C_TI_{00, 00}=A_H^2C_T\bigg(1-\frac{1}{d}\bigg). 
\eea 
\\ 

\noindent 
Now we show a regularization method for extracting KS metric
in CFT$_2$ and CFT$_3$. 
The 2-point function of $H_{\mathrm{mod}}$ is given by: 
\bea
&&
\langle H_{\mathrm{mod}}(x_1, x_2)H_{\mathrm{mod}}(x_3, x_4)\rangle
\nn\\
&=&\frac{c}{72}z^2{}_2F_1(2, 2, ; 4, z)+\frac{c}{72}\bar{z}^2{}_2F_1(2, 2, ; 4, \bar{z})
\nn\\
&=&\frac{c}{72}\frac{6\big((z-2)\ln(1-z)-2z\big)}{z}+\frac{c}{72}\frac{6\big((\bar{z}-2)\ln(1-\bar{z})-2\bar{z}\big)}{\bar{z}}, 
\eea 
where $c$ is a central charge. 
Now we do an expansion as the following: 
\bea
x_1= x_4+\delta x; \qquad 
x_3=x_2+\delta y. 
\eea 
Therefore, we obtain that: 
\bea
u&\equiv&\frac{x_{12}^2x_{34}^2}{x_{13}^2x_{24}^2}=z\bar{z}
\nn\\
&=&\frac{\big(x_{42}^2+2(x_{42}\cdot \delta x)+(\delta x)^2\big)
\big(x_{24}^2+2x_{24}\cdot\delta y+(\delta y)^2\big)}
{\big(x_{42}^2+2x_{42}\cdot(\delta x-\delta y)+(\delta x-\delta y)^2\big)
x_{24}^2
}
\nn\\
&=&
1+\frac{2}{x_{42}^2}
\bigg(\eta_{\mu\nu}-\frac{2(x_{42})_{\mu}(x_{42})_{\nu}}{x_{42}^2}\bigg)\delta x^{\mu}\delta y^{\nu}
+\cdots; 
\nn\\ 
v&=&\frac{x_{14}^2x_{23}^2}{x_{13}^2x_{24}^2}=(1-z)(1-\bar{z})\rightarrow 0, 
\label{uv}
\eea 
where $x_{jk}\equiv x_j-x_k$.   
Now we introduce $z$ as that $z\equiv 1+\epsilon-\tilde{\epsilon}$, 
where $1\gg \epsilon/\tilde{\epsilon}\gg \epsilon$. 
Up to the first-order term of $\epsilon$ and $\tilde{\epsilon}$, the $\bar{z}$ is one.
The $\tilde{\epsilon}$ is an additional regularization parameter to avoid the divergence. 
Therefore, we choose $\epsilon$ as that 
\bea
\epsilon=
\frac{2}{x_{42}^2}
\bigg(\eta_{\mu\nu}-\frac{2(x_{42})_{\mu}(x_{42})_{\nu}}{x_{42}^2}\bigg)\delta x^{\mu}\delta y^{\nu}
+\cdots.
\eea
We can observe the following universal term from $\langle 0| H_{\mathrm{mod}}(x+\delta x, y)H_{\mathrm{mod}}(y+\delta y, x)|0\rangle$, 
\bea
\langle 0|\partial_{\mu, x} H_{\mathrm{mod}}\partial_{\nu, y} H_{\mathrm{mod}}|0\rangle\delta x^{\mu}\delta y^{\nu}\sim  \frac{c}{6}\epsilon\ln(\tilde{\epsilon}). 
\eea 
Now we do the identification as that: $x_1=x+\delta x$; $x_2=y$; $x_3=y+\delta y$; $x_4=x$. 
Therefore, the symmetric part of QMGT is given by: 
\bea
ds^2
&=&\frac{\langle 0|\partial_{\mu, x} H_{\mathrm{mod}}\partial_{\nu, y} H_{\mathrm{mod}}|0\rangle
}{(2\pi i)^2}\delta x^{\mu}\delta y^{\nu}
\sim-\frac{1}{4\pi^2}\frac{c}{6}\epsilon\ln(\tilde{\epsilon})
\nn\\
&=&
\frac{1}{4\pi^2}a_2\ln(\tilde{\epsilon})
\frac{4}{x_{42}^2}
\bigg(-\eta_{\mu\nu}+\frac{2(x_{42})_{\mu}(x_{42})_{\nu}}{x_{42}^2}\bigg)\delta x^{\mu}\delta y^{\nu}, 
\eea 
where $a_2=c/12$.  
\\

\noindent
The vacuum state is also the eigenstate of the modular Hamiltonian for the spherical case. 
Therefore, we can directly use the vacuum state to calculate QMFT. 
Hence we obtain KS metric from the symmetric part of QMGT. 
For the CFT$_1$ and other even-dimensional CFTs, we can use the same regularization method to obtain KS. 
\\

\noindent 
For CFT$_3$, the 2-point function of modular Hamiltonian is ($G_{\Delta, l}$ is a global conformal block of conformal dimension $\Delta$ and spin $l$, whose explicit form is obtained from the recursion relation): 
\bea
\langle 0| H_{\mathrm{mod}}(x_1, x_2)H_{\mathrm{mod}}(x_3, x_4)|0\rangle=C_1G_{3, 2}
=\frac{\pi}{2} a_3\frac{z(z+4\sqrt{1-z}-8)-8\sqrt{1-z}+8}{z\sqrt{1-z}}. 
\eea 
When we consider the same expansion of $\epsilon$ and $\tilde{\epsilon}$,  
the 2-point function of modular Hamiltonian for the linear term of $\epsilon$ becomes 
\bea
\langle 0| H_{\mathrm{mod}}(x+\delta x, y)H_{\mathrm{mod}}(y+\delta y, x)|0\rangle
\sim
\frac{\pi}{4} a_3
\bigg(\frac{\epsilon }{2 {\tilde \xe}^{3/2}}-12 \sqrt{\tilde \xe} \epsilon -\frac{7 \epsilon }{2\sqrt{\tilde \xe}}+8 \epsilon\bigg).
\eea 
Because we have $\bar{z}=1$ up to the first-order term of $\epsilon$ and $\tilde{\epsilon}$, it is equivalent to doing a direct expansion of $\epsilon$ and $\bar{\epsilon}$ from $C_1G_{3, 2}/2$.  
Ignoring the divergent terms, the symmetric part of QMGT also gives the KS metric
\bea
ds^2\sim\frac{1}{4\pi^2}\pi a_3\frac{4}{x_{42}^2}\bigg(-\eta_{\mu\nu}+\frac{2(x_{42})_{\mu}(x_{42})_{\nu}}{x_{42}^2}\bigg)\delta x^{\mu}\delta y^{\nu}. 
\eea
One can use the same regularization in odd-dimensional CFTs, except for $d=1$. 
Hence we have the regularization method for all dimensions. 
Because we always take limit $z\rightarrow 1$ for extracting KS metric, we only need the recursion relation of the conformal block $G_{\Delta, l}$ at $z=\bar{z}$ \cite{Dolan:2011dv}, 
\bea
&&
(l+d-3)(2\Delta+2-d)G_{\Delta, l}(z)
\nn\\
&=&(d-2)(\Delta+l-1)G_{\Delta, l-2}(z)+\frac{2-z}{2z}(2l+d-4)(\Delta-d+2)G_{\Delta+1, l-1}(z) 
\nn\\
&&
-\frac{\Delta(2l+d-4)(\Delta+2-d)(\Delta+3-d)(\Delta-l-d+4)^2}
{16(\Delta+1-\frac{d}{2})(\Delta-\frac{d}{2}+2)(l-\Delta+d-5)(l-\Delta+d-3)}
G_{\Delta+2, l-2}(z)\,.\quad 
\eea  

\noindent 
The boundary condition $G_{\Delta, 1}$ and $G_{\Delta, 0}$ are known in terms of generalized hypergeometric functions. 
After solving for $G_{d, 2}$, one can do the regularization to obtain KS metric from the symmetric part of QMGT.  
The spacetime interval for $d>1$, in general, is given by
\bea
ds^2\sim\frac{1}{4\pi^2}(-1)^{\frac{d}{2}-1}a_d\ln(\tilde{\epsilon})\frac{4}{x_{42}^2}
\bigg(-\eta_{\mu\nu}+\frac{2(x_{42})_{\mu}(x_{42})_{\nu}}{x_{42}^2}\bigg)\delta x^{\mu}\delta y^{\nu}
\eea 
for the even-dimensional CFT; 
\bea
ds^2\sim\frac{1}{4\pi^2}(-1)^{\frac{d+1}{2}}\pi a_d\frac{4}{x_{42}^2}
\bigg(-\eta_{\mu\nu}+\frac{2(x_{42})_{\mu}(x_{42})_{\nu}}{x_{42}^2}\bigg)\delta x^{\mu}\delta y^{\nu}
\eea
for the odd-dimensional CFT. 
When $d=1$, the spacetime interval is that
\bea
ds^2\sim\frac{1}{4\pi^2}a_1\frac{4}{x_{42}^2}
\bigg(-\eta_{\mu\nu}+\frac{2(x_{42})_{\mu}(x_{42})_{\nu}}{x_{42}^2}\bigg)\delta x^{\mu}\delta y^{\nu}, 
\eea 
where $a_1\equiv a_2/2$. 

\section{Wilson Line}
\label{sec:3}
\noindent 
Here we want to establish the relation between KS metric \cite{deBoer:2016pqk} and Wilson lines correlator in the gauge formulation of 3d Einstein gravity.  
One can expand the logarithm of the expectation value of Wilson lines correlator as in the following:  
\bea
&&
\ln \langle 0| W(x_1, x_2)W(x_3, x_4)|0\rangle 
\nn\\ 
&\equiv&
\ln\big(\langle 0| W(x_1, x_2)\rangle\langle W(x_3, x_4)|0\rangle+ \langle 0| W(x_1, x_2)W(x_3, x_4)|0\rangle_{\mathrm{connected}}\big)
\nn\\ 
&=&
\ln\big(\langle 0|W(x_1, x_2)|0\rangle \big)+\ln\big(\langle 0|W(x_3, x_4)|0\rangle\big)
+\frac{\langle 0|W(x_1, x_2)W(x_3, x_4)|0\rangle_{\mathrm{connected}}}{\langle 0|W(x_1, x_2)|0\rangle\langle 0|W(x_3, x_4)|0\rangle}
\nn\\
&&+{\cal O}\bigg(\frac{1}{c}\bigg), 
\eea
where $W(x_1, x_2)$ is a Wilson line with two ending points, $x_1$ and $x_2$, and
$\langle W(x_1, x_2)W(x_3, x_4)\rangle_{\mathrm{connected}}$ is the connected part of the Wilson lines correlator. 
In the AdS$_3$ Einstein gravity theory, one realized the following identity \cite{Huang:2019nfm, Huang:2020tjl}
\bea
S_{EE}=\lim_{n\rightarrow 1}\frac{1}{1-n}\ln\langle 0| W_{\cal R}|0\rangle, 
\eea
where $S_{EE}$ is entanglement entropy of CFT$_2$, and $W_{\cal R}$ is an $n$-sheet Wilson line operator.  
It is easy to show that 
\bea
\ln(\langle 0| W_{\cal R}|0\rangle)=\frac{c}{6}(n-1)\ln(\langle 0 |W|0\rangle )+\cdots, 
\eea
where $\cdots$ does not contribute to entanglement entropy under the one-sheet limit $n\rightarrow 1$.   
Hence we obtain that
\bea
S_{EE}=-\frac{c}{6}\ln\langle 0| W|0\rangle. 
\eea 
The Wilson line operator, in general, depends on the conformal dimension $h$. 
Using the result of entanglement entropy can fix the coefficient of $h$: 
\bea
-\frac{c}{6}\ln\langle 0|W(x_1, x_2)|0\rangle=\frac{ch}{3}\ln(x_1-x_2)^2=\frac{c}{3}\ln|x_1-x_2|. 
\eea
Therefore, we obtain $h=1/2$. 
We can also include the quantum correction of Wilson lines by the identification $c=\tilde{c}+26$ \cite{Huang:2019nfm, Huang:2020tjl}, where $\tilde{c}=3l/(2G_3)$, $l\equiv 1/\sqrt{-\Lambda}$, and $G_3$ is 3d gravitational constant.  
The $\Lambda$ is a cosmological constant. 
Hence the one-loop correction of bulk gravity theory does not modify our result. 
\\

\noindent 
Now we define a new quantity from the connected correlator of Wilson lines: 
\bea
I\lbrack A: B\rbrack
&=&
S_{EE}\lbrack A\rbrack+S_{EE}\lbrack B\rbrack+\frac{c}{6}\ln\langle 0|W(x_1, x_2)W(x_3, x_4)|0\rangle
\nn\\
&=&
\frac{c}{6}\frac{\langle 0|W(x_1, x_2)W(x_3, x_4)|0\rangle_{\mathrm{connected}}}{\langle 0| W(x_1, x_2) 0|\rangle\langle 0| W(x_3, x_4)|0\rangle}+{\cal O}(c^0),   
\eea
in which the ending points of the region $A$ are $x_1$ and $x_2$ and the region $B$ are $x_3$ and $x_4$. 
From the relation between Wilson line and the identity Virasoro OPE block \cite{Fitzpatrick:2016mtp}, we obtain the connected correlator
\bea
\label{connectedWilson}
\frac{c}{6}\frac{\langle W(x_1, x_2)W(x_3, x_4)\rangle_{\mathrm{connected}}}{\langle W(x_1, x_2)\rangle\langle W(x_3, x_4)\rangle}
=\frac{1}{12}z^2{}_2F_1(2, 2; 4, z)+\frac{1}{12}\bar{z}^2{}_2F_1(2, 2; 4, \bar{z})+{\cal O}\bigg(\frac{1}{c}\bigg).  
\label{connectedWilson}
\nn\\
\eea 
Hence we obtain the relations between $I$ or the correlator of Wilson lines and the two-points of $H_{\mathrm{mod}}$
\bea
\frac{c}{6}I\lbrack (x_1, x_2): (x_3, x_4)\rbrack
=\langle H_{\mathrm{mod}}(x_1, x_2)H_{\mathrm{mod}}(x_3, x_4)\rangle
+{\cal O}(c^0).
\eea 
\\

\noindent 
From the computation of QMGT, the KS metric is given by:
\bea
&&
ds^2
\nn\\ 
&\propto&-\bigg\langle 0\bigg|\frac{\partial H_{\mathrm{mod}}}{\partial x}\frac{\partial H_{\mathrm{mod}}}{\partial y}\delta x\delta y\bigg|0\bigg\rangle 
\nn\\
&\sim& 
\langle 0|H_{\mathrm{mod}}(x-\delta x, y)H_{\mathrm{mod}}(x, y+\delta y)|0\rangle 
-\langle 0|H_{\mathrm{mod}}(x-\delta x, y)H_{\mathrm{mod}}(x, y)|0\rangle
\nn\\ 
&&
-\langle 0|H_{\mathrm{mod}}(x, y)H_{\mathrm{mod}}(x, y+\delta y)|0\rangle
+\langle 0|H_{\mathrm{mod}}(x, y)H_{\mathrm{mod}}(x, y)|0\rangle. 
\eea
We can rewrite the kinematic space in terms of correlators of Wilson lines on a given time slice:  
\bea
&&
ds^2
\nn\\ 
&\sim&
\frac{c}{6}\big(
I(L_1\cup L_2: L_2\cup L_3)
-I(L_1\cup L_2: L_2)
-I(L_2: L_2\cup L_3) 
+I(L_2: L_2)
\big)
\nn\\ 
&=&
\frac{c^2}{36}\bigg(\ln\big(\langle 0| W(x-\delta x, y)W(x, y+\delta y)|0\rangle\big)
-\ln\big(\langle 0|W(x-\delta x, y)W(x, y)|0\rangle\big)
\nn\\
&&
-\ln\big(\langle 0|W(x, y)W(x, y+\delta y)|0\rangle\big) 
+\ln\big(\langle 0|W(x, y)W(x, y)|0\rangle\big)\bigg),
\eea
where 
\bea
L_1\equiv (x-\delta x, x), \ 
L_2\equiv (x, y), \ 
L_3\equiv (y, y+\delta y). 
\eea 
\\

\noindent 
Because the KS metric on a time slice is also given by the second-order derivative of a geodesic line \cite{Czech:2015qta}: 
\bea
ds^2&=&\frac{\partial^2 S_{EE}(x, y)}{\partial x\partial y}\delta x\delta y
\nn\\ 
&\sim& 
S_{EE}(x-\delta x, y)+S_{EE}(x, y+\delta y)-S_{EE}(x, y)-S_{EE}(x-\delta x, y+\delta y)
\nn\\ 
&=&
S_{EE}(L_1\cup L_2)+S_{EE}(L_2\cup L_3)-S_{EE}(L_2)-S_{EE}(L_1\cup L_2\cup L_3). 
\eea 
Combining the above results, we obtain the following equivalence: 
\bea
&&
S_{EE}(L_1\cup L_2)+S_{EE}(L_2\cup L_3)-S_{EE}(L_2)-S_{EE}(L_1\cup L_2\cup L_3)
\nn\\
&\sim&\frac{c^2}{36}\bigg(\ln\big(\langle 0| W(x-\delta x, y)W(x, y+\delta y)|0\rangle\big)
-\ln\big(\langle 0|W(x-\delta x, y)W(x, y)|0\rangle\big)
\nn\\
&&
-\ln\big(\langle 0|W(x, y)W(x, y+\delta y)|0\rangle\big) 
+\ln\big(\langle 0|W(x, y)W(x, y)|0\rangle\big)\bigg)
\nn\\ 
&=& 
\frac{c}{6}\big(
I(L_1\cup L_2: L_2\cup L_3)
-I(L_1\cup L_2: L_2)
-I(L_2: L_2\cup L_3) 
+I(L_2: L_2)
\big). 
\eea 
We can observe that the above equality still holds if one does the following replacement 
\bea
\frac{c}{6}I(A: B)\rightarrow I_M(A: B)\equiv S_{EE}(A)+S_{EE}(B)-S_{EE}(A\cup B), 
\eea 
where $I_M$ is the mutual information of the regions $A$ and $B$. 
The result is non-trivial because we do not assume that two intervals are far from each other and also include the quantum correction of Wilson lines. 
Our derivation only guarantees the equivalence up to the second-order of $\delta x$ and $\delta y$. 
Note that the connected Wilson line correlators do not vanish for two disjoint intervals with a far separation. 
Hence our result at least holds up to the one-loop correction for $c$ (beyond the minimum surface). 

\section{Outlook}
\label{sec:4}
\noindent 
We showed that QMGT is a combination of KS metric (symmetric part) \cite{Czech:2016xec, deBoer:2016pqk} and MBC (anti-symmetric part) for a spherical case in all CFTs. 
Therefore, we realized the geometrization of CFTs, similar to the story of QGT \cite{Provost:1980nc}. 
Now one can obtain Quantum Kinematic Space by calculating the two-points of $H_{\mathrm{mod}}$ without relying on conformal symmetry. 
\\ 

\noindent 
In CFT$_1$ and CFT$_2$, we can compute the KS metric for an excited state using the connected 2-point correlator of the identity Virasoro OPE block. 
On a vacuum state, this correlator reduces to the 2-point function of modular Hamiltonian \eqref{connectedWilson}. 
The correlator becomes the identity Virasoro block in an excited state. 
The connected part is the global block of stress tensor after a coordinate transformation. 
It is not difficult to show that the resulting KS metric agrees with the known result.
\\

\noindent 
We formulated the quantum relation between the connected correlator of Wilson lines and QMGT in the gauge formulation of 3d Einstein gravity. 
This relation provides a new Quantum Information interpretation to the connected correlator of Wilson lines as the mutual information. 
However, the derivation is only valid up to a one-loop quantum correction for $c$ and second-order expansion for $\delta x$ and $\delta y$. 
Hence verifying the following generalized form in the 3d Einstein gravity should teach us more about Quantum Gravity from Quantum Information: 
\bea
&&
\frac{c^2}{36}\big(\ln(\langle W_{L_1\cup L_2}W_{L_2\cup L_3}\rangle) 
-\ln(\langle W_{L_1\cup L_2}W_{L_2}\rangle) 
-\ln(\langle W_{L_2}W_{L_2\cup L_3}\rangle) 
+\ln(\langle W_{L_2}W_{L_2}\rangle)\big) 
\nn\\ 
&=&
I_M(L_1\cup L_2: L_2\cup L_3)
-I_M(L_1\cup L_2: L_2)
-I_M(L_2: L_2\cup L_3) 
+I_M(L_2: L_2)
\nn\\
&=&
S_{EE}(L_1\cup L_2)+S_{EE}(L_2\cup L_3)-S_{EE}(L_2)-S_{EE}(L_1\cup L_2\cup L_3).
\eea 

\section*{Acknowledgments}
\noindent
Chen-Te Ma would like to thank Nan-Peng Ma for his encouragement.
\\

\noindent
Xing Huang acknowledges the support of the NSFC Grants No. 11947301 and No. 12047502. 
Chen-Te Ma acknowledges the YST Program of the APCTP;  
China Postdoctoral Science Foundation, Postdoctoral General Funding: Second Class (Grant No. 2019M652926); 
Foreign Young Talents Program (Grant No. QN20200230017).


\end{document}